\documentstyle[preprint,revtex]{aps}

\begin{document}
\draft

\begin{title}
Monte Carlo methods for the nuclear  shell  model
\end{title}

\author{C.~W.~Johnson, S.~E.~Koonin, G.~H.~Lang, and  W.~E.~Ormand}
\begin{instit}
W. K. Kellogg Radiation Laboratory \\
California Institute of Technology \\
Pasadena, CA 91125, USA \\
\end{instit}
\begin{abstract}
We present novel Monte Carlo methods for treating the interacting shell model
that allow exact calculations much larger than those heretofore possible. The
two-body interaction is linearized by an auxiliary field;  Monte Carlo
evaluation of the resulting functional integral gives ground-state or thermal
expectation values of few-body operators.  The ``sign problem'' generic to
quantum Monte Carlo calculations is absent in a number of cases.  We discuss
the favorable scaling of these methods with nucleon
number and basis size and their suitability to parallel
computation.
\end{abstract}


\narrowtext

The shell model (valence fermions confined by a one-body potential and
influencing each other through a residual two-body interaction) is a ubiquitous
framework for the quantum many-body problem and is often the method of choice
for describing nuclear structure.\cite{ref1}  For example, the low-lying
spectra and
one-body transition matrix elements of nuclei with $17\leq A\leq 39$ are very
well described by the exact diagonalization of an effective two-body
Hamiltonian within the single-particle basis of the
$1d_{5/2}$-$1d_{3/2}$-$2s_{1/2}$ orbitals.\cite{ref2}

Unfortunately, the combinatorial growth of the dimension of the many-body basis
with both the number of valence particles ($A$) and the size of the
single-particle basis~($N$)
precludes an exact treatment for most larger nuclei and forces what are often
{\it ad~hoc} truncations of the many-body basis.  The $J^\pi=0^+$, $T=0$ states
of $\rm
{}^{28}Si$ are obtained by constructing and diagonalizing an $839\times 839$
Hamiltonian matrix in the $sd$-shell basis noted above.  A similar calculation
for $\rm {}^{60}Zn$ in the middle of the next major shell (10 neutrons and 10
protons in the $1f_{7/2}$-$1f_{5/2}$-$2p_{3/2}$-$2p_{1/2}$ orbitals) would
increase
the dimension to $5,053,574$ and is clearly beyond the reach of
today's computers.

In this Letter, we discuss Monte Carlo methods for the exact treatment of a
shell-model hamiltonian, $H$.  They are based on using the imaginary-time,
many-body evolution operator, $\exp(-\beta H)$, either to define the
canonical or grand canonical ensemble at a temperature $\beta^{-1}$ or, for
large $\beta$, to filter a many-body trial state  to the
exact ground state; thus both thermodynamic and ground-state properties can be
obtained.  Relative to direct diagonalization, these methods scale much more
gently with $N$ and/or~$A$, and hence hold the promise of
extending complete shell-model calculations to much larger systems.

Our methods broadly follow previous work on the Hubbard model \cite{ref3} and
coordinate-space fermion systems,\cite{ref4} but differ significantly in detail
due
to the peculiarities of the nuclear shell model. A general expression
for $H$ is
\begin{equation}
H=\sum_\alpha\epsilon_\alpha O_\alpha+{1\over2}\sum_\alpha
V_\alpha O_\alpha^2\eqnum{1}
\end{equation}
 where the $O_\alpha$ are a set of non-commuting
one-body operators, the $c$-numbers $\epsilon_\alpha$ are related to the
single-particle energies, and the $c$-numbers $V_\alpha$ characterize the
residual interaction. A minimal choice for the $O_\alpha$ is the set of
hermitian and anti-hermitian parts of the multipole density operators,
$\rho_{\rm KM}=[a^\dagger\times\tilde a]_{\rm KM}$.  In this case, the
$V_\alpha$ are related by a Pandya transformation to the usual two-body
matrix elements of the residual interaction. However, even for a fixed
$H$, there is considerable freedom in writing Eq.~(1).  For example, arbitrary
symmetrized two-body matrix elements of the residual interaction have no effect
on the
(antisymmetric) eigenstates and eigenvalues of $H$, but do change
the~$V_\alpha$. Additionally, the density operators could be supplemented by
the
suitably hermitized pair operators $\Delta^\dagger_{\rm JM}=[a^\dagger\times
a^\dagger]_{\rm JM}$ and their adjoints, which might be convenient given the
strong pairing character of the residual interaction.

For a concrete illustration, we consider refining a trial determinant
$\vert\Phi\rangle$ to the
exact ground state; details of implementation and other applications will be
given elsewhere.\cite{ref5}   For any observable operator $B$, we define:
\begin{equation}
\langle B\rangle={\langle\Phi\vert e^{-\beta H/2}Be^{-\beta
H/2}\vert\Phi\rangle
\over\langle\Phi\vert e^{-\beta H}\vert\Phi\rangle}\eqnum{2}
\end{equation}
In the limit $\beta\rightarrow\infty$, $\langle B\rangle$ approaches the
required ground state expectation value (as long as $\vert\Phi\rangle$ is not
orthogonal to the true ground state).  We divide the ``time'' $\beta$ into an
even number, $N_t$, of equal intervals of duration $\Delta\beta=\beta/N_t$ and
introduce a real auxiliary field $\sigma_{\alpha n}$ coupled to each operator
$O_\alpha$ at each time slice $n=1,\ldots, N_t$.  Then, in the limit
$\Delta\beta\rightarrow0$,  the Hubbard-Stratonovich trans

formation allows $\langle B\rangle$ to be written as
\begin{equation}
\langle B\rangle ={\int d\sigma W(\sigma)B(\sigma)\over
\int d\sigma W(\sigma)D(\sigma)}\, .\eqnum{3}
\end{equation}
Here, $d\sigma=\displaystyle{\prod_{\alpha n}} d\sigma_{\alpha n}$ and the
integral extends over all real fields.  The positive definite ``weight
function'' is
\begin{equation}
W(\sigma)=\exp\left(-{1\over2}\Delta\beta\sum_{\alpha n}\vert V_\alpha\vert
\sigma^2_{\alpha n}\right)\bigl\vert\langle\Phi\vert
U(N_t,0)\vert\Phi\rangle\bigl\vert
\, .\eqnum{4}
\end{equation}
Here, the one-body evolution operators are
$U(k,j)\equiv\displaystyle{\prod^k_{n=j+1}}
U_n$, with $U_n=\exp(-\Delta\beta h_n)$, the one-body Hamiltonian being
\begin{equation}
h_n=\sum_\alpha\epsilon_\alpha O_\alpha+\sum_\alpha V_\alpha s_\alpha
\sigma_{\alpha n}O_\alpha\eqnum{5}
\end{equation}
where $s_\alpha=+1$ if $V_\alpha<0$ and $s_\alpha=i$ if $V_\alpha>0$.  The
remaining functions in the integrand are the ``observable''
\begin{equation}
B(\sigma)\equiv{{\rm Re}\langle\Phi\vert
U(N_t,N_t/2)BU(N_t/2,0)\vert\Phi\rangle\over
\bigl\vert\langle\Phi\vert U(N_t,0)\vert\Phi\rangle\bigl\vert}\eqnum{6a}
\end{equation}
and the ``sign,''
\begin{equation}
D(\sigma)\equiv{{\rm Re}\langle\Phi\vert U(N_t,0)\vert\Phi\rangle\over
\bigl\vert\langle\Phi\vert U(N_t,0)\vert\Phi\rangle\bigl\vert}\, .\eqnum{6b}
\end{equation}
Similar expressions, valid for $N_t$ either even or odd, can be
written for the grand canonical trace (with the introduction of
chemical potentials) or for the canonical trace (using coherent states or an
expansion in the fugacity).\cite{ref5}

The advantages of these expressions are manifest.  The $h_n$ are
one-body operators, so that the action of $U_n$ on a determinant is
particularly simple.  Indeed, the $O_\alpha$, $h_n$, and $U_n$ can be
represented as $N\times N$ matrices,
while the determinant  $\vert\Phi\rangle$ can be defined by $A$ $N$-dimensional
single-particle wavefunctions.   Hence all of the quantum mechanics
can be  done explicitly by matrix manipulation, while the integrals
over
the $\sigma_{\alpha n}$ can be done by standard Monte Carlo methods
(e.g., the Metropolis algorithm).  Of course, $N_t$ must be chosen large enough
so that the $\Delta\beta\rightarrow0$ limit is practically satisfied.

As noted above, our methods have several antecedents.  Relative to simulations
of the Hubbard model,\cite{ref3}  the number of single-particle orbitals is
considerably
smaller than the typical number of lattice sites, and even relatively small
fillings are of interest.  Further, the shell-model operators $O_\alpha$ are
fully non-local in the basis, rather than being at most nearest-neighbor
hopping.  Finally, proportionately smaller values of $\beta$ are of interest in
the nuclear problem.  As shown below, $\beta\vert V_\alpha\vert
{\mathrel{\raise.3ex\hbox{$<$}\mkern-14mu
             \lower0.6ex\hbox{$\sim$}}} 20$ can
yield useful results, while for an on-site Hubbard repulsion of 1~eV, a
temperature of 100~K implies $\beta\vert V_\alpha\vert\sim 120$.  Other
authors have presented path-integral calculations of the nuclear shell
model.\cite{ref6}
 However, these have generally been based  on the uncontrolled Static Path
Approximation~(SPA) in which the fields have no time-dependence
(harmonic corrections to the SPA have also been considered). Further, only
the simplest schematic interactions have been treated.

To demonstrate our methods, we calculate ground-state properties of ${}^{24}$Mg
and ${}^{48}$Cr via Eqs.~(3--6), respectively in the $sd$- and $pf$-shell
bases.  While these calculations are not quite as demanding as those at the
middle of the respective shells (${}^{28}$Si and ${}^{60}$Zn), these nuclei
are deformed and hence present non-trivial problems.  As noted above,
Hamiltonians
in the $sd$-shell can be diagonalized exactly, allowing a rigorous test
of our methods; diagonalization of the ${}^{48}$Cr Hamiltonian in the
$pf$-space is at today's computational limit.

For these first calculations, we have used an interaction of the ``pairing $+$
multipole'' form:
\begin{equation}
V=V_{\rm pair}+V_{\rm multi}\, .\eqnum{7}
\end{equation}
This interaction is complicated enough to demonstrate the power of our methods,
but is not meant to produce observables that can be compared directly to
experiment.  The pairing is taken to be of constant strength and isovector
monopole character:
\begin{equation}
V_{\rm pair}=-G\sum_{jj^\prime} \biggl[a^\dagger_j\times
a^\dagger_j\biggr]_{J=0~T=1}\cdot
\biggl[\tilde a_{j^\prime}\times\tilde a_{j^\prime}\biggr]_{J=0~T=1}\,
.\eqnum{8}
\end{equation}
For the multipole force, we take
\begin{equation}
V_{\rm multi}=\sum_{L=0,2,4}C_L f_L(r_1)f_L(r_2)Y_L(\hat r_1)\cdot Y_L(\hat
r_2) \eqnum{9}
\end{equation}
where the radial functions are $f_0=1$ and
\begin{equation}
f_{L>0}(r)=r^{L-1}{dg\over dr}\, ,\eqnum{10}
\end{equation}
with $g$ a Woods-Saxon function.  Two-body matrix elements of $V_{\rm multi}$
were calculated in an harmonic oscillator single-particle basis.  The values of
the interaction parameters and single-particle energies were chosen to roughly
reproduce the $T=1$ two-body matrix elements of the $sd$-shell Wildenthal
interaction.\cite{ref2}

We note that our interaction is not as simple as it might appear.
As we rewrite the pairing interaction in the form of Eq.~(1)
with the $O_\alpha$ chosen as {\it density} operators,
we must introduce all possible
fields.  This contrasts sharply with previous treatments of the
nuclear shell model,\cite{ref6}  where the forces were chosen so that only a
small number of fields (5~in the case of SPA for a quadrupole-quadrupole
interaction) were needed.  In our largest calculations, 9600~fields were
required.

Figs. 1(a,b) show, for $^{24}$Mg and
$^{48}$Cr, respectively, $\langle H\rangle$ and $\langle J^2\rangle$ as
functions of $\beta$ for various $\Delta\beta$.  The $^{24}$Mg trial state was
taken to be the prolate Hartree solution, while for $^{48}$Cr we chose the
maximally prolate trial state in only the $1f_{7/2}$ orbital.  There is a clear
relaxation with increasing $\beta$ and a
convergence as $N_t$ is increased (or $\Delta\beta$ decreased) at fixed
$\beta$.

At large $\beta$, $\langle H\rangle$ for ${}^{24}$Mg is in excellent agreement
with the results of
direct diagonalization, given by the solid lines. For $^{48}$Cr in Fig.~1(b),
we
cannot diagonalize the  Hamiltonian in the full model space; however one can
compare with $\langle H\rangle$ calculated at two levels of truncation of the
model space ($f^8_{7/2}$ and $f^8_{7/2}+2p-2h$ excitations)  and note that our
result is below these two.  We therefore believe that we have successfully
calculated the
ground state of $^{48}$Cr.  To our knowledge, no calculation has ever before
treated this nucleus in the full $pf$-shell.  We can also, in principle,
calculate any $n$-body observable of the
ground state.  Note that for both nuclei, $\langle J^2\rangle$ is large for
small $\beta$ (because of the deformed trial state), but vanishes for large
$\beta$ (because of the $J=0$ ground state), and that the SPA is quite
inadequate.

Using the methods described in Ref.~5, we can also calculate observables in the
canonical ensemble.  Results for ${}^{20}$Ne are shown in Fig.~1(c), where they
are also compared with direct diagonalization.  Results for ${}^{24}$Mg, which
cannot be obtained by other methods, are shown in Fig.~1(d).

An important point is the relatively gentle scaling of our methods
with $A$ and/or $N$, as illustrated in Table~1. The time to construct
the Hamiltonian matrix and find the lowest-lying eigenvalues using
the Lanczos algorithm increases rapidly with the dimension
of the many-body basis, and memory requirements also
grow dramatically with the basis size.  The well-known OXBASH code
\cite{ref7} works well in the $sd$-shell, but would find ${}^{48}$Cr extremely
difficult. A full $pf$-shell calculation of ${}^{48}$Cr may be possible
with the Glasgow-Los Alamos-Seattle code,\cite{ref8} but would likely require
in excess of 2Gb mass storage.  ${}^{60}$Zn
would require in excess of 1000 times more memory and, from that
consideration alone, is presently  impossible to treat exactly by
diagonalization in the full $pf$-shell.

In contrast, Monte Carlo calculations scale with the number of auxiliary
fields to be integrated over, which is, in turn, the square of the number of
single-particle orbits (the number of possible density operators), 144 and 400
for the $sd$- and $pf$-shells, respectively. (The six-fold increase in time
from the $sd$- to the $pf$-shell is due not only to this factor, but also to
the longer
correlation times in the Metropolis Monte Carlo sampling.)
Furthermore, because the Monte Carlo
integration involves statistically independent samples, it is well-suited to
parallel computer architectures. The Monte Carlo calculations presented here
were performed on the Intel Touchstone Gamma and Delta parallel supercomputers,
with 64 and 512 nodes respectively.  Each of these  nodes is an i860 processor
that runs our code at  a speed of 5 double-precision Mflops.  The number of
node-hours
for each calculation is given in the third column of Table~1. ${}^{60}$Zn
calculations are clearly feasible (indeed, we have done them) and even a full
$sd$-$pf$ calculation would be possible.

We have found that when a general residual interaction (e.g., the
antisymmetrized two-body matrix elements of the Wildenthal
interaction \cite{ref2}) is arbitrarily linearized via the Hubbard-Stratonovich
representation, the so-called ``sign problem'' can arise; that is, $D(\sigma)$
in Eq.~(6b) can be both positive and
negative, so that the resulting cancellation leads to unacceptable
statistical fluctuations.  However, it is possible to prove that an interaction
of the form~(7-9), when applied to $T_Z=0$ nuclei, is completely
free of the sign problem.

We have been able to show empirically that the flexibility in formulating the
integral to be calculated (using varying proportions of the density or pairing
breakups, and adding `non-physical' symmetrized two-body matrix elements) can
be used to mitigate the sign problem.  While we can
use this flexibility only in an {\it ad hoc} manner at present and do not
fully understand the sources of (and solutions to) the sign problem for a
general interaction,
we are optimistic  that a wide range of
interactions will be tractable.

Future work will include not only a further investigation of the
sign problem and use of more general residual interactions,
but also applications to partition functions and level
densities, strength functions (e.g.,~the E1, E2 and
Gamow-Teller responses), high-spin nuclei, and
problems in nuclear astrophysics.

\acknowledgments
We are grateful for discussions with other members of the Caltech
nuclear theory group.  C.~W.~Johnson and W.~E.~Ormand
acknowledge Caltech Divisional and DuBridge postdoctoral fellowships,
respectively.  This work was supported in part by the National
Science Foundation  (Grant Nos. PHY90-13248 and PHY91-15574).

\figure{$\langle H\rangle$ and $\langle J^2\rangle$ for selected $sd$ and $pf$
cases as functions of $\beta$.  $\diamondsuit - {\rm SPA}$;
${\vcenter{\vbox{\hrule height.4pt\hbox{\vrule width .4pt height 6pt\kern 6pt
\vrule width .4pt}\hrule height .4pt}}} -\Delta\beta=0.5$,
$\Delta-\Delta\beta=0.25$, $\circ-\Delta\beta=0.125$, $\vbox{\hrule width 6pt
height 6pt }-\Delta\beta=0.0625~{\rm MeV}^{-1}$.  (a) ${}^{24}$Mg ground state;
(b) ${}^{48}$Cr ground state; (c) ${}^{20}$Ne, canonical ensemble; (d) ${}^{24}

$Mg, canonical ensemble.  Solid lines indicate the results of direct
diagonalization, except for (b), where $\langle H\rangle$ for two different
truncations of the model space are shown.  Where not shown, error bars are
smaller than the size of the plotting symbols.  Each point corresponds to
between 2000 and 5000 samples.}


\begin{table}
\caption{Computational time for shell model calculations. }
\begin{tabular}{cccc}
Nucleus & Basis Size$^a$ &\multicolumn{2}{c}{Time (i860 node hours)}\\
($N/A$)& $J=0,T=0$ & Diagonalization$^d$ & Monte Carlo$^e$\\
\multicolumn{4}{l}{\rule{3.38truein}{2truept}}\\
${}^{24}$Mg & 325 & .04$^b$ & 607\\
(24/8)& (28,503) \\
\tableline
${}^{28}$Si & 839 & 0.4$^b$ & 670  \\
(24/12) & (93,710) \\
\tableline
${}^{48}$Cr & 9,741 & $\sim40^c$& 3353 \\
(40/8) & (1,963,461) \\
\tableline
${}^{60}$Zn & 5,053,574 & ${\sim4\times10^6}~{}^c$ & 3558 \\
(40/20) & ($2.5\times 10^9$) \\
\end{tabular}
\end{table}

\noindent
{$^a$}~~In parentheses are the total number of $J_Z=0$ $m$-scheme Slater
determinants from which $J=0,T=0$ states are projected.

\noindent
{$^b$}~~Scaled from OXBASH calculations on VAX 3100 workstation.

\noindent
{$^c$}~~Estimate using the Glasgow-Los Alamos-Seattle code.\cite{ref8}

\noindent
{$^d$}~~Time to compute only the ground state.

\noindent
{$^e$}~~Time for 3000 samples, giving the precision shown in Fig.~1, with
$N_t=24$ and with adequate decorrelation of the Metropolis random walk.  No
extraordinary effort has been made to optimize performance.

\end{document}